# Conception and Use of Ontologies for Indexing and Searching by Semantic Contents of Video Courses

**Merzougui Ghalia[1], Djoudi Mahieddine[2] and Behaz Amel[3]**

**[1] Departement of computer science, Faculty of Science, Batna University, (05000) Algeria**

**[2] Laboratory XLIM-SIC and IRMA a Research Group, UFR Sciences SP2MI, University of Poitiers Teleport 2, Boulevard Marie et Pierre Curie BP 30179 86962 Futuroscope, Chasseneuil Cedex- France**

**[3] Departement of math, Faculty of Science, Batna University, (05000) Algeria**

**Abstract**

Nowadays, the video documents like educational courses available on the web increases significantly. However, the information retrieval systems today can not return to the users (students or teachers) of parts of those videos that meet their exact needs expressed by a query consisting of semantic information. In this paper, we present a model of pedagogical knowledge of current videos. This knowledge is used throughout the process of indexing and semantic search segments instructional videos. Our experimental results show that the proposed approach is promising.

***Keywords:*** *video course, ontology, OWL, conceptual indexing, semantic search, vector method adapted.*

## 1. Introduction

The e-Learning is largely based on multimedia materials and particularly on videos. Many institutes, schools and associations on the web diffuse video lectures on scientific conferences, seminars and thesis dissertations or habilitations (e.g. INRIA, ENS, Aristotle…). Some Universities (or virtual campus) diffuse on the Internet their lectures as audio or video (are cited as an example: MIT, Berkeley, Strasbourg, MedNet and Lausanne). In addition, university lectures are grouped in thematic portals such as WebTV Lyon3 or SciVee (one of many examples of sites dedicated to science videos). These videos are recorded in different formats: i.e. video streamed (or podcast) or structured multimedia documents (where video and presenter's voice are synchronized with slides), and this for a live broadcast or delayed.

While these video documents are more accessible to their richness and semantic expressiveness and their numbers are growing more and more, their treatment remains problematic. In particular, the search for relevant video sequences according to criteria related to the semantic content is not trivial. This can affect the learner while revising it or the researcher (or teacher) who wants to reuse a portion of a video for him. It is often more convenient for a user (learner or teacher) to use semantic information in its query (scientific concepts) to get the most relevant answers. Therefore a process of indexing and searching by the semantics of this type of video should be set up.

Before reaching this stage, it should be noted that it is virtually impossible to achieve the semantic level, starting from a low-level analysis of video content. Interpretations of the contents of a video, which are semantically richer, make the task of the indexer more complicated than the case of a by keyword indexing. This is because he must choose the best index to describe content very rich in information. One meets the same difficulty in the research process. So we must first develop models capable of describing and modeling the semantic content of these videos in order to facilitate access, reuse and navigation by semantics.

In this context, the processing of video content using techniques of knowledgebase is an interesting idea. In the perspective of Semantic Web, which is becoming a basis for distance learning environments, the ontology provides a rich semantic better than any other method of knowledge representation [1]. In a teaching platform, the precision of a search for educational content can be improved if based on the conceptual vocabulary defined in ontology while avoiding the ambiguities in terminology and allowing inferences that reduce noise and increase relevance.

Our work aims to develop ontological models to form a conceptual vocabulary shared between the teachers and learners. We will use this vocabulary in the annotation of videos from university lectures. Then, we seek to develop a system for indexing and searching the semantic content of video segments, based on their ontological annotation to overcome the lack of such a tool actually.
In Section 2, we present different approaches to video indexing, namely, classical / semantic, automatic / semi-automatic, low level / high level. Section 3 will come later, and will present some work using ontologies for indexing documents in the two areas of interest, namely the e-learning and audiovisual. Then we describe in Section 4 our approach which is divided into five stages: ontological modeling of semantic content of video courses, their annotation based on models developed, conceptual indexing, conceptual research and finally experimentation. We conclude by specifying the limits and prospects of our approach.

## 2. Semantic Indexation of Video

The indexing of the video document is difficult and complicated. This type of material does not decompose to easily identifiable units as is the case for text document. It is therefore necessary to have tools able to segment, to describe and to annotate the content [2]; this is the task of annotating video document.
Charhad has defined video annotation as the process by which text informations (or other) are associated with specific segments of video material to enrich the content. This information does not modify the document but is just mapped to it. The annotation is often considered to be a laborious task that requires human intervention; however, it remains in high demand for describing the semantic content of a video.
Several standards are used to describe or to annotate multimedia content, such as Dublin Core [3] and MPEG7 [4], using a defined list of attributes such as creation date, authors, image resolution, etc... Dublin Core is used to describe the data cataloging bibliographic records. MPEG 7 is used to describe, in a fine low-level, visual and sound elements of an audiovisual document (such as texture, dominant color ...). But Troncy in [5] found that the descriptors of the latter are too low to accommodate all the needs of semantic description. These standards are far from satisfactory because each provides few mechanisms of knowledge. Therefore ontologies can supplement them in order to access the level of multimedia semantics.
Indexing based on ontologies to represent the documentation granules is called semantic indexing. It consists in choosing the set of concepts and instances of ontology as a representation language of the documents. The granules are then indexed by concepts that reflect their meaning rather than words quite often ambiguous. One should use an ontology reflecting domain or domains of knowledge discussed in the document collection [6].
Hernandez identified two steps to semantic indexing [6]. The first step is to identify the concepts or instances of the ontology in the granules also called conceptual annotation. The second step is to weight the concepts for each document based on the conceptual structure from which it originates.
In what follows, we will cite some works that use ontologies for indexing video documents corresponding to two areas: e-learning and broadcasting.

## 3. Existing Works

The use of ontology in the context of indexing has grown in recent years in various fields; we cite two that interest us: the audiovisual sector [2], [7] and [8] and the field of e-Learning [9], [10] and [11]. In this latter, several studies are based on the general idea of indexing document fragments on the basis of different kinds of ontologies: i.e. ontology document structure, domain ontology or pedagogical ontology (figure, formula, equation...), to reuse them to compose more or less automatically new resources.
For example, the IMAT project [9] is turned to the use of ontology for indexing. The handling of documents requires the use of an ontology called 'document' which adds to the traditional domain and pedagogical ontologies. The ontology of the course is divided into three sections describing the content, the context and the structure. The content is the domain ontology and the other two parts are related to the pedagogical aspects (structuring the chapters, nature of the parties, etc.).
The project MEMORAe [10] describes organizational memory training based on two ontologies. The first is domain ontology which describes the concepts of the training: individual (student, teacher...), documents (book, web page ...), educational activities (courses, TP...). The second is an application ontology that specifies all the concepts useful for specific training such as algorithms or statistics.
The project Trial Solution [11] consists in taking each educational resource and breaks it down into learning objects 'OP'. Each OP is represented by its semantic content and its relationship with the other OP and metadata that concern. An annotation tool was developed by the project that indexes each node by metadata and by the terms of a thesaurus. May be mentioned other similar works in the field of e-learning [12], [13] and [14].

In the audiovisual field, we cite the work of [7] which articulates a specific conceptual knowledge of a domain through ontology in the context of indexing of audiovisual materials on the theme 'Sport TV emission' In this work, there are two ontologies:

- An audio-visual ontology to standardize the meaning of terms commonly used to describe the structure and format of audiovisual materials. For example a schema that indicates that a sports magazine is like sports program and that always begins with start sequence plateau, followed by a number of sequences that are either plateau sequence or a sequence launch-plateau-report and ends with a sequence end.
- The second is the domain ontology that models the concepts of a particular sport which is his example cycling (tour de France, sports magazine, etc.)...

The work of Isaac [8] is based on the semantic description of the content of television programs on the theme of medicine. It combines multiple ontologies namely: ontology of AV and thematic ontologies related medical fields (MENELA which describes the field of coronary disease and includes concepts related to cardiac surgery correspondent to the topic of the corpus, GALEN contains concepts related to all medical fields).

In [2], Charhad proposed a model for the representation of semantic content of videos. This model allows synthetic and integrated consideration of information components (image, text, sound). He developed a number of tools to extract concepts (name of a person, a geographical place or an organization) and one for detection and recognition of the identity of the speaker which is based on the analysis of automatic transcription of speech in a video.

We also cite some recent studies, [15] and [16], on the treatment of the semantic content of video representation of the field of e-learning. Dong et al. [15], offer a model of multi-ontology annotation of multimedia documents. But they focus in their paper on video presentations of lectures, seminars and corporate training. Each segment is annotated from a Multimedia Ontology (OM) and several domain ontologies. The 'OM' ontology is based on the standard MPEG7, but focuses on the aspect of content description. It contains three types of classes or concepts: multimedia concepts (image, video, audio, video segment, etc…), non-multimedia concepts (agent, place, time, etc…) and descriptor concepts of domain ontologies (such as Gene Ontology 'GO '). We note that the pedagogical aspect is missing in the annotation in this work.

First, we must state that in e-learning, the course material is available on the web in two categories of documents: static multimedia documents (such as web page, pdf, doc, etc…) and dynamic or temporal documents (such as video, audio or SMIL [17]). The number of documents of the second category continues to grow while indexing jobs in the field of e-learning mentioned above are based only on documents of the first type.

Second, the temporal nature of such documents creates a number of constraints on their management. Indeed, the specificity of these documents is to be temporal objects. Inherently this temporality does not present itself and can not be stored and this has several consequences. One of them is the imposition of rate of reading the document, if the video is an hour, it takes an hour to see it and if the information sought begins at the 12th minute and lasts 10 minutes, then you have to wait all this time or scroll through the first 11 minutes to find her.

Third, the video courses posted on the web are annotated in general metadata (format, creation date, author, title, keywords, and sometimes abstract). Note that the use of free text (keywords, abstract), to describe the content, prevents the control of the description's semantic and this severely limits the possibilities of reasoning.

Search engines currently available do not allow a search by the semantic content of a sequence (or segment) in a video course because it is not logically structured, nor semantically described. There are no tools that offer this possibility so far.

So our contribution is in the context of offering the community of e-learning (learner or teacher) a system that helps to search the semantic content of instructional video segments.

## 4. Approch

In a context of video courses information seeking by semantic content, modeling is an important and necessary task from which the index will be formulated and the by which research process will be more efficient and more accurate. Our approach comes as a modeling and indexing of pedagogical video courses and research through the semantics of the segments in such videos.

At the theoretical level, our contribution consists in the proposal and the construction of two types of ontologies, one for the pedagogical structuring of a video course and the other for describing the semantic content of its various granules. Both ontologies will be used in the phase of conceptual annotation.

At the experimental level, our contribution consists, first, in the conceptual annotation of a corpus of video course about continuous professional training broadcast on the Web from the University NETTUNO under the project

MedNet'U. Through this project MedNet'U (Mediterranean Network of Universities), satellite channels Rai Nettuno Sat forwarded academic lessons on professional training arguments in four languages: Italian, French, English and Arabic.

The annotation is done on ten video lessons from the course 'data structure and algorithm' (in French). Then, our contribution consists in the implementation of the prototype IRSeCoV: a system of indexing and semantic searching of pedagogical video
segments through conceptual annotations associated with the corpus and we follow by an experiment.

## 4.1 Construction of Ontologies

We need two ontologies to model the content of courses in video format. The first will be built for pedagogical structuring of a video course and will be called pedagogical ontology of the video course. The second is the ontology of the domain of teaching and, as its name suggests, it will model the knowledge of a subject area (a teaching modulus) for a deeper semantic description of this type of course. We begin by describing the latter.

### 4.1.1 Ontology of the Domain of Teaching

A domain or area of teaching is a single module within training. A module addresses or teaches one or more concepts. A concept can be broken down into several concepts; it may depend on one or more concepts as may be the prerequisite of one or more concepts as well. So, three types of relationships may exist between two concepts: '*is_decomposed_into*', '*depends*', and '*is_prerequisite*'. Note that '*is_prerequiste*' has the characteristic of transitivity while '*depends*' is symmetrical and '*is_decomposed_into*' is anti-symmetric.
It is noted that the exploitation of the characteristics of these relationships can generate or infer instances not found in the basis of the original facts.
Consider the example of teaching domain "*data structure*" which discusses the concepts: function, parameter, parameter passing by value, list, pointer and recording. Instances of relationships that can exist between these concepts are represented as follows:

- *is_decomposed* (function, parameter). The function is composed of parameter.
- *depends* (function, parameter_passing_by_value). Sinc e this relationship is symmetrical, the inference system can deduce the next instance:
- *depends* (parameter_passing_by_value, function).
- *is_prerequiste* (pointer, list) and
- *is_prerequiste* (list, tree) =>

*is_prerequiste* (pointer, tree). The relation is transitive.
Fig. 1 shows this ontology in the form class diagram.

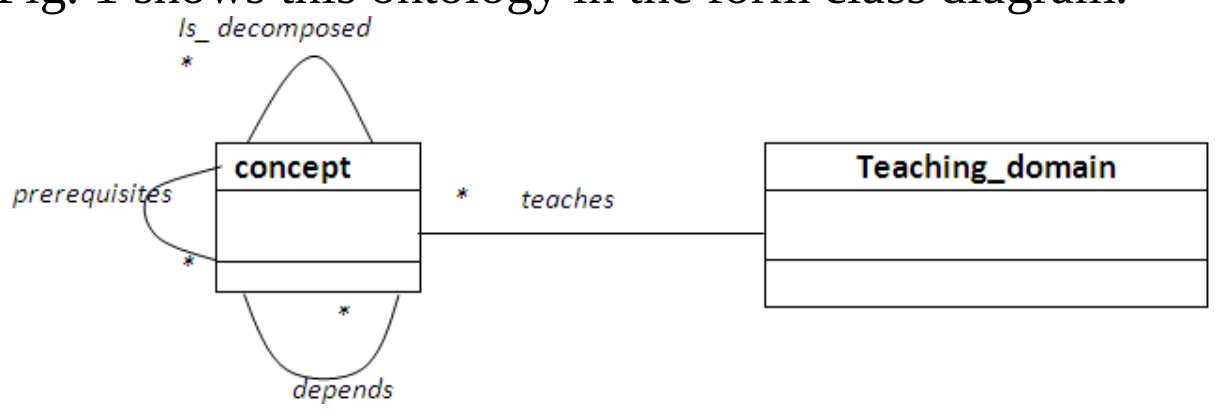

Fig. 1. Domain Ontology of Teaching.

We can mention the existence of instances of the class concepts that are identical. We cite as an example: loop and repetition’s instruction, address and pointer, record and structure, two-dimensional table and matrix, parameter and attribute, etc... This semantics is specified in OWL [18] by the property 'sameAs' between individuals.
An instance of this ontology creates domain ontology to teach D (a specific module). It can be manual or semi-automatically. In the second case, this process makes use of language engineering tools (such as LEXTER) for the extraction of candidate terms from one or more textual course materials in a particular field D. These terms represent the extension of the class concept and should be selected, sorted by an expert in the field and organized hierarchically according to the relation '*is_decomposed_into*’. Then, the semantics of the domain is refined by the precision of identical instances of the class concept and the relations of the two bodies '*is_prerequisites*' and '*depends*' that can exist between different concepts.
For our part, we manually created an ontology for the module ‘*Data Structure’* with the publisher '***protégé 2000***'. Below one can find an excerpt from the OWL code generated by this tool. We used the French language because the video lessons we annotate are in French.

```
<?xml version="1.0"?>
<rdf:RDF
    xmlns=http://www.owl-
ontologies.com/Ontology1277939276.owl#
        ………
<Teaching_domaine rdf:ID="structure_de_donnee">
<teachs>
 <concept rdf:ID="instruction">
  <is_decomposed rdf:resource="#affectation"/>
  < is_decomposed
   rdf:resource="#instruction_de_controle"/>
 <is_decomposed
  rdf:ID="#instruction_de_repetition"/>
 <concept rdf:ID="boucle"/>
   <owl:sameAs rdf:resource=
    "#instruction_de_repetition"/>
 </concept>
 <concept rdf:ID="passage_parametre_par_valeur">
   <depends rdf:resource="#fonction"/>
 </concept>
 <concept rdf:ID="pointeur">
   <prerequisites rdf:resource="#liste"/>
 </concept>
```

```
…
</teachs>
…
```

### 4.1.2 Pedagogical Ontology of a Video course

A video course is presented in one or more video lessons. Each video lesson (it can be a chapter or sub chapter) is divided or segmented into several temporal segments. The segment corresponds to the explanation of one or more slides with the same title. So the segment in this case must represent an idea or a subject or unit of interest which will be returned by our search system.

A segment or a slide contains one or more pedagogical objects 'POb' (or Learning Objects). It can be a definition, an example, an exercise, a solution_exercice, an illustration, a rule, a theorem, a demonstration, etc....

While viewing some video lessons from the corpus that we chose, we found that a slide contains a definition of a concept followed by a small example. We also noticed that one example can be presented in two or three slides with the same title; we opted for this manner of structuring segments.

This POb concerns one or more concepts of a teaching domain. The relation '*concerns*' manages the alignment of two ontologies: i.e. pedagogical ontology of the video course (POV) and domain ontology of teaching (DOT). To do this, there is a need to import the second (DOT) into the first (POV) (see Fig. 2).

A question: why a POb concerns a number of concepts and not one. If we take the example of the video lesson with the title 'functions', where there is a slide with an exercise on the use of a table as a parameter of a function, we see that the POb-type 'exercise' concerns the three concepts (underlined) of the domain ontology of teaching ‘Data Structure’ (to our knowledge, there is no search engine that responds to a request of this type).

## 4.2 Annotation Process

Some authors, such as Charhad, consider or call the annotation phase as assisted or manual indexing. This phase describes the pedagogical video documents by considering two aspects: one is pedagogical and identifies the components of the educational structure of the document (slide, learning object type definition, example ...) and the second is thematic and describes each element as a concept in the field.

Troncy and Isaac used the tool SegmentTool, while Charhad used VidéoAnnex. In our case, we have developed a new tool for segmentation and annotation of video course called OntoCoV and based on ontologies we created.

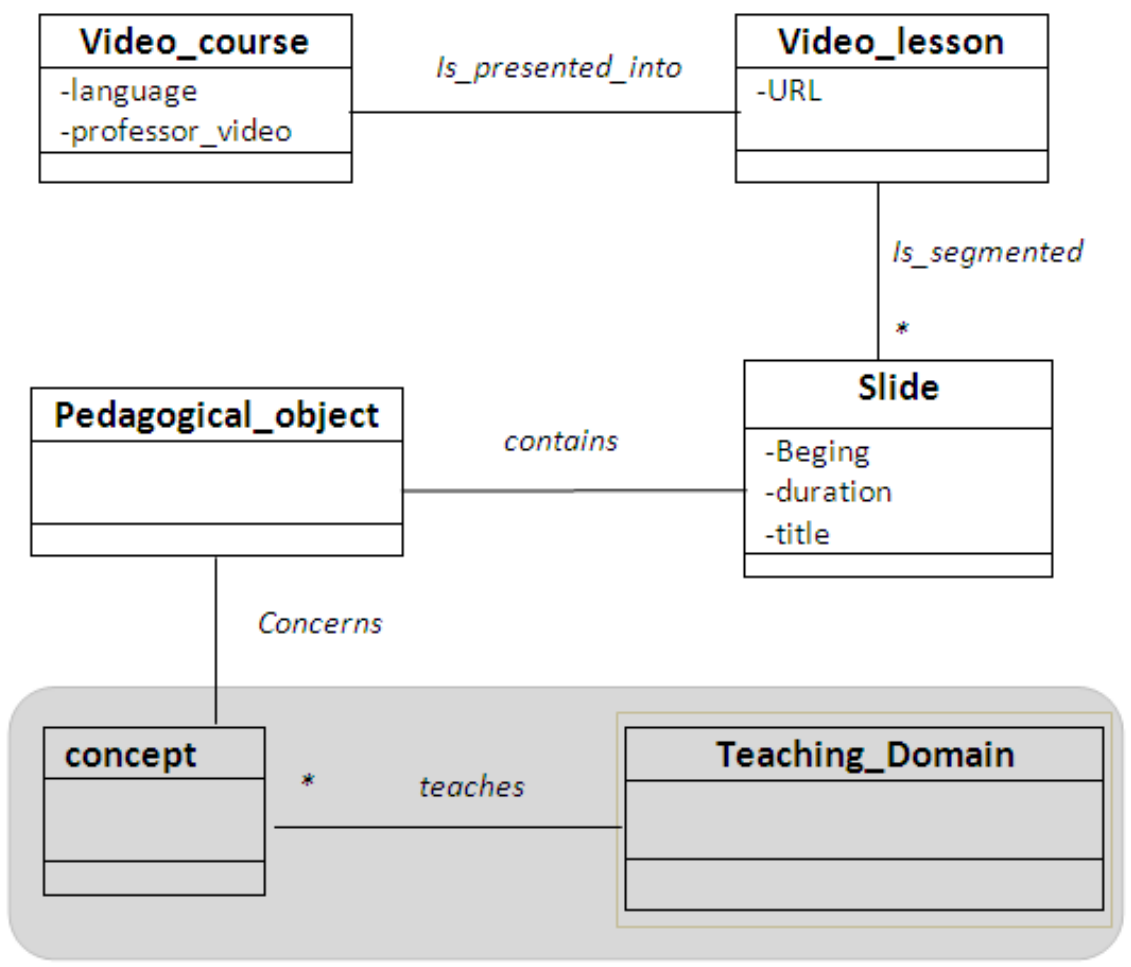

Fig. 2. The Pedagogical Ontology of the video course.

The description or annotation via OntoCoV begins with the localization or temporal segmentation into entities regarding an indivisible concept. It comes to identify segments in time, where each corresponds to exposure of one or more slides with the same title. Then, each segment must characterize its pedagogical structure. These two steps generate the instantiation of the ontology of video course. Next, a description of the semantic content of each segment is made by the association of concepts of the ontology of a teaching field that is particular to different pedagogical objects (POb).

This ontology must have a close relationship with the video lesson being annotated. So OntoCoV gives its user the possibility to integrate ontology of a particular area. This ontology is presented in our tool as a tree graph, which allows the user to quickly browse and select, at all levels (hierarchy of concepts), the concept that seems pertinent for its indexing. At the end, the system will generate all the annotations in the operational language OWL. These annotations provide a basis of facts that will be exploited in subsequent phases.

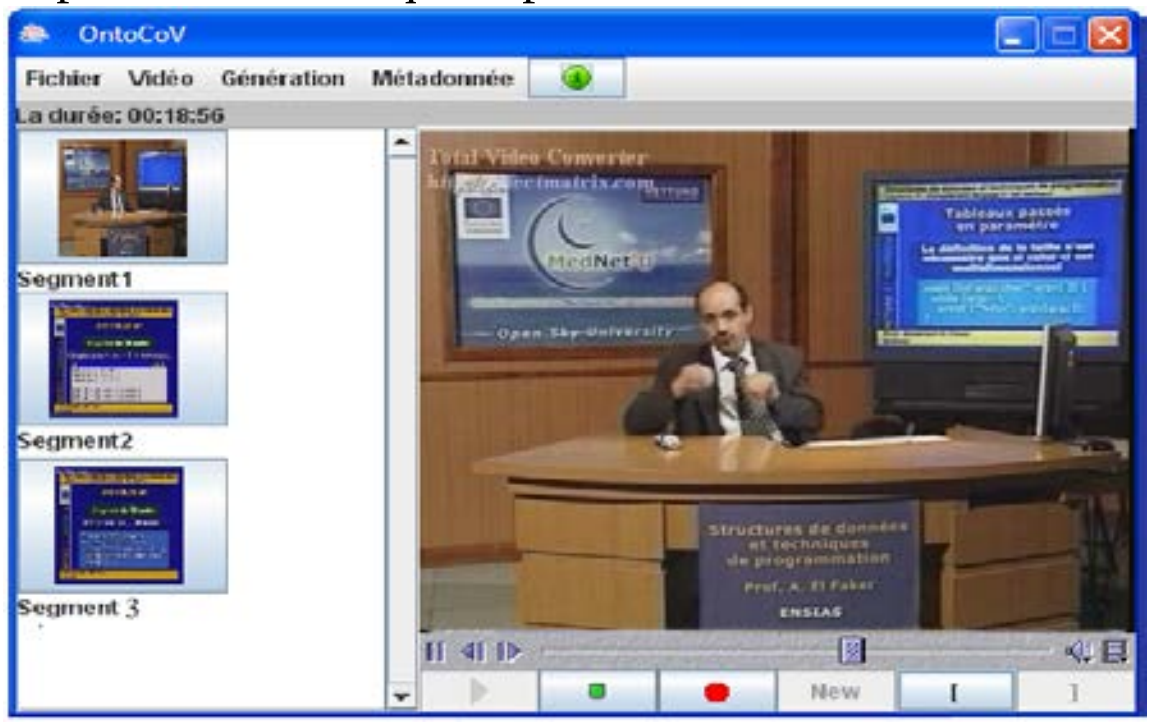

Fig. 3. The interface of OntoCoV tool.

Fig. 3 shows the interface of the tool OntoCoV which is divided into two regions:

(a) Area to watch the video with buttons to play the video, stop it, create a segment, etc...
(b) Region to display all the segments slide built.

A small separate window appears by clicking on a segment of the area (b). It contains a list of pedagogical objects forming the selected segment. We can describe each POb by associating a list of concepts of the ontology of teaching field already built into the tool.

After the annotation of a video lesson of the course 'data structure and programming techniques', the tool will generate the following OWL code:

```
…
<video_course rdf:ID="structure_de_donnee">
 <is_presented_into>
  <lesson_video rdf:ID="fonction">
         <URL
rdf:datatype="http://www.w3...#string">
        http://.../fonction.wmv </URL>
   <is_segmented rdf:resource="#slide_2"/>
   <is_segmented rdf:resource="#slide_3"/>
   <is_segmented rdf:resource="#slide_7"/>
  </lesson_video>
 </is_presented_into>
<langage rdf:datatype="&xsd#time">frensh</langage>
…
</cours_video>
<slide rdf:ID="slide_2">
 <Duration rdf:datatype="&xsd#time">00:03:22
  </Duration>
 <Beging rdf:datatype="&xsd#time">00:02:01
  </Beging>
 <Title rdf:datatype="&xsd#string">introduction au
  function</Title>
 <contains>
  <POb rdf:ID="definition_1">
     <concerne rdf:resource="&p1#adresse"/>
     <concerne rdf:resource="&p1#fonction"/>
  </POb>
 </contains>
 <contains>
  <POb rdf:ID="exemple_1">
   <concerne rdf:resource="&p1#valeur_retournee"/>
   <rdfs:comment rdf:datatype="&xsd#string">
    differents type de valeurs retournée
   </rdfs:comment>
  </POb >
 </contains>
</slide>
…
```

## 4.3 Conceptual Indexing

Once the concepts of both ontologies have been identified in the temporal segments, we move to the phase concept’s weighting for each slide.

In this part of our work, we present an index structure that can pose queries on temporal segments of video document. For this we use the vector model of Salton [19] while adjusting the calculation of the weight TF_IDF (Term Frequency_Inverse document Frequency) for our needs, drawing on the works of [20] and [21].

It is proposed that the document for a video lesson is no longer represented by a vector but a matrix of concepts and temporal segments. Since the segments are described by concepts instead of words, we compute the weight of concepts with respect to segments in which they appear. Thus we define the new formula CF_ISDF (Concept Frequency_Inverse Segment and Document Frequency), as follows:

$$CF-ISDF(c,s,d)=CF(c,s,d)\times$$
$$\times ISF(c,s,d)\times IDF(c,d)$$
$$ISF(c,s,d)=\log\left(\frac{S_d}{SegF(c,s)}\right)$$
$$IDF(c,d)=\log\left(\frac{S_d}{SegF(c,s)}\right) \qquad (1)$$

$CF(c,s,d)$: The number of occurrences of concept *c* in the segment *s* of the document *d*.

$D$ : Set of all documents (video lessons) of the corpus.

$S_d$ : Number of segments in the document *d*.

$SegF(c,s)$ : Number of segments in the document *d* in which the concept *c* appears.

$DF(c)$ : Number of documents containing the concept.

This formula allows us to balance the concept not only by its frequency in the segment *s* on a document *d*, but also its distribution in the document (*ISF*(*c*, *s*, *d*)); this last measure represents the discriminatory strength of a concept *c* in the document *d*. The distribution of the concept in the corpus (*IDF*(*c*, *d*)) is also important. If a concept appears in several documents, it is less representative for a given document with respect to another concept that appears only in only one document. It is the discriminatory strength in the corpus.

## 4.4 Conceptual Search

Now we come to the search phase which is also called the interrogation phase. It includes:

- formulating the need for information through query,
- translating the query into an internal representation defined by a query template,
- comparing the request to document’s indexes in the corpus by the correspondence function,
- presenting the results in order of relevance.

Formulation of the query: The corpus of video courses covers several subjects or areas of teaching. For each TD we associate an ontology according to the model developed above. These ontologies, which were used during the annotation, will be used to help the user formulating its query. Our system provides an interface for visualization and exploration of an ontology of a particular TD, chosen by the user, to guide him to browse the tree of this ontology and giving him the opportunity to choose the concepts of his query (see Fig. 5).

The reason that led us to choose this way of query formulation is twofold:

- the user has indeed difficulty to specify his need and to express it,
- one must remove ambiguities and improve the precision and recall of our system.

Query template: For each query (as for segments), we associate a vector. We can assign a weight to the concepts of the query. We assign the value 1 when the concept is present and 0 otherwise.

Correspondence Function: We adapt a measure of pertinence from classical vector model. The relevance of a query *Q* over a segment *S* of the document *D* is:

$$pertinence(S_D, Q) = \cos(V_{S,D}, V_Q) \quad (2)$$

$V_{S,D}$ and $V_Q$ are respectively the vectors of weight of concepts in segment *S* of document *D* and query *Q*.

The Search Results: is a list of references to the most pertinent segments, viewed in order of pertinence in a page coded in HTML + time. Hence the user can see and read a selected video segment on the same page.

## 4.5 Prototype and Experiment

To evaluate our approach, we implemented a prototype called IRSeCoV (Abbreviation of French translation of: Indexing and Semantic Search in Video Course). Our system aims to allow a more accurate and relevant search of pedagogical video segment. It has several components (see Fig. 4 and Fig. 5) which allow it to be modular.

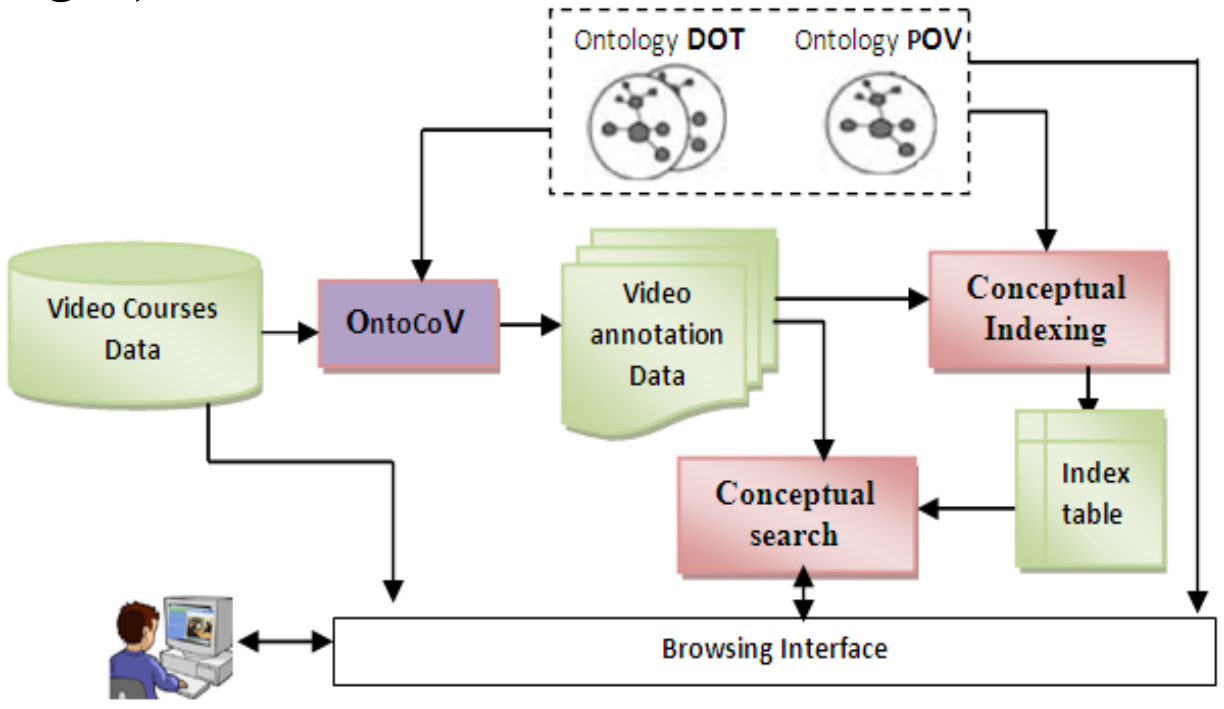

Fig. 4. The General architecture of IRSeCoV system.

To evaluate our system, we have two approaches:

- The first concerns the evaluation of the structure of the index. It comes to calculate the time of indexing, the storage space of the index relative to the size of the corpus, the time of constructing the ontology and the time of corpus annotation. Calculating of time of constructing the index does not assess the value of the index.
- The second concerns the evaluation of the relevance of the index by testing its impact on research using traditional measures of relevance (recall and precision).

Initially we tested the weight of some concepts taken from the index table that was generated by our system.

The experimentation of our system was done on a corpus (annotated by OntoCoV) of 9 video lessons (from 25) of the module 'data structure and programming techniques' which was given during a continuing professional formation by the virtual university NETTUNO under the project MedNet'U. The following table shows some concepts and the list of segments in which they appear. The segment is defined by the document number (video lesson) and the segment number (slide) in this document.

Table 1: List of concepts associated with segments

| Concepts | List of segments |
|---|---|
| *Pointeur* | {($D_1$,$S_7$), ($D_4$,$S_1$), ($D_5$,$S_5$), ($D_5$,$S_{12}$), ($D_6$,$S_2$), ($D_8$,$S_9$), ($D_8$,$S_{14}$), ($D_9$,$S_4$)} |
| *Parametre_formel* | {($D_4$,$S_2$)} |

*CF-ISDF*(Pointeur,$S_2$,$D_6$) = 1×log(13/2) ×log(9/6) = 0,7589.

*CF-ISDF*(Pointeur,$S_9$,$D_8$) = 6×log(16/11) ×log(9/6) = 0,9110.

*CF-ISDF*(Pointeur,$S_{14}$,$D_8$)= 1×log(16/11) ×log(9/6) = 0,3038.

*CF-ISDF*(Parametre_f,$S_2$,$D_4$) = 1×log(14/2) ×log(9/1) = 4,2756.

- If a concept appears in two segments of the same lesson, the frequency determines the most pertinent segment. See the concept 'pointeur' in the two segments ($D_8$,$S_9$) and ($D_8$,$S_{14}$).
- If a concept appears in segments from different lessons, then the discriminatory strength ISF determines the most pertinent segment. See the concept 'pointeur' in the segments ($D_6$,$S_2$) and ($D_8$,$S_{14}$).

If a concept appears in one or two segments at most in the corpus, it will have a great weight because of the discriminatory value IDF; as can be seen with the concept ‘parametre_formel’.

The user can specify in his query the pedagogical object with the concepts he seeks. The system returns a list of segments sorted by pertinence. For each segment, it displays the name of the lesson, its beginning, its duration, its title and more importantly, the pedagogical objects included in the segment with a comment. The user can thus select the segments as needed (see Fig. 5).

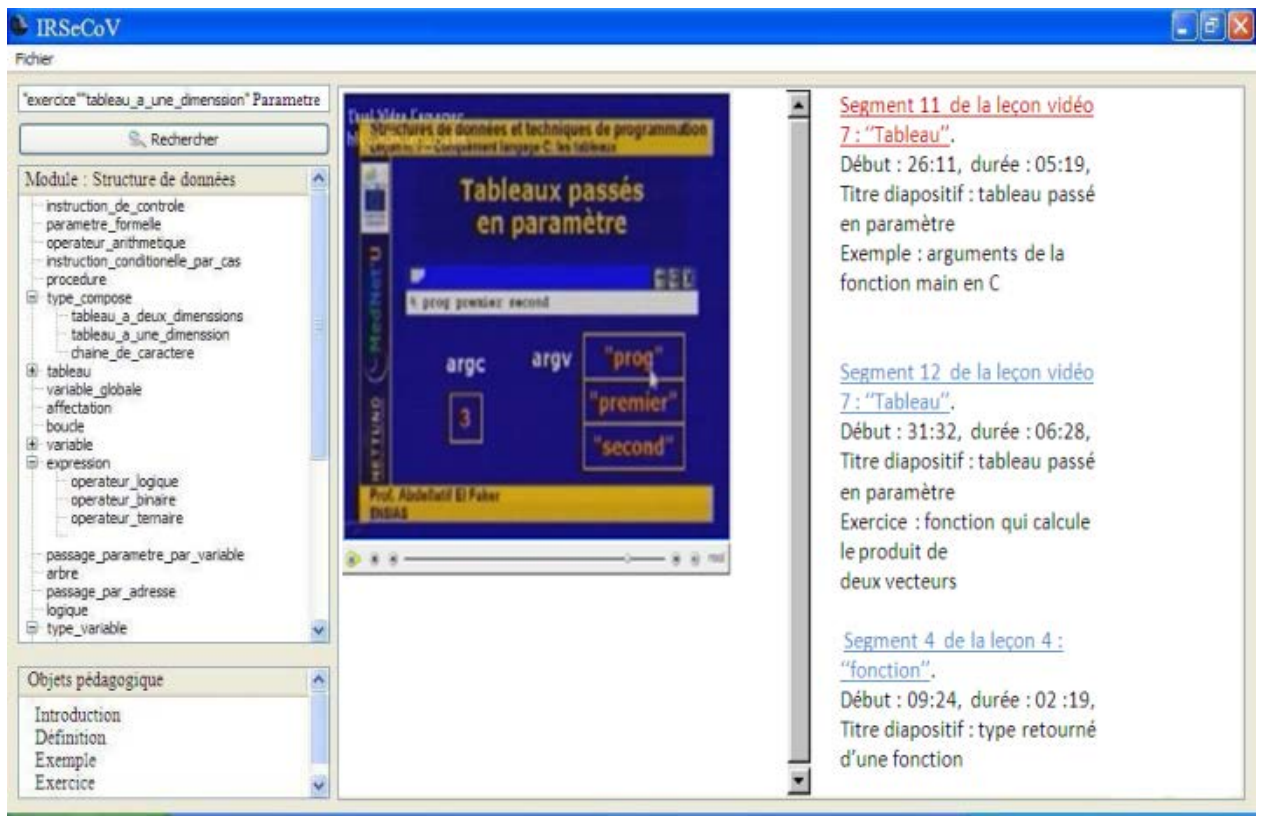

Fig. 5. System interface IRSeCoV.

We plan to expand the testing of our system on a corpus of video courses from different teaching field to assess its relevance by calculating recall and precision.

## 5. Conclusions

We have presented in this paper an approach of searching by the semantic content of pedagogical video segments using ontologies. We built two ontologies, the first structure, pedagogically, a video course and the second models the knowledge of a teaching field.

We realized a new tool called OntoCoV that generates the annotation of video lesson in OWL-based on ontologies.

Then we detail the indexing and the conceptual searching of all annotated video course by adapting the vector method. We have defined a new formula CF-ISDF to calculate the weight of a concept in a video segment. To implement this approach, we developed the prototype IRSeCoV and we experimented this system on a few video lessons annotates on the module 'data structure'. The obtained results show the feasibility and benefits of using ontologies to search by the semantic content in pedagogical video segments.

However, it is important to note that our approach is far from being finished and that it has to evolve in the near future.

To improve the research relevance, we think to use semantic inference in the search of content. The results (explicit assertions) returned by the conceptual search may be supplemented by implicit assertions derived or inferred from the knowledge base by exploiting the semantic relations between concepts (e.g. transitivity, similarity, etc…)

We suggest also extending the ontological model, by integrating knowledge about the profile of learners to guide our system to the adaptation of video segments based on their profiles.

**Acknowledgments**

G. Merzougui would like to thank a lot both M. Moumni and A. Behloul for discussions, comments and suggestions that have greatly enriched the work.

## References

[1] V. Psyché, Olavo, J. Bourdeau, Apport de l'ingénierie ontologique au environnements de formation à distance, International Review sticef : science et technologie de l'information et de la communication pour l'Education et la formation, Vol 10, 2003.

[2] M. Charhad, Modèles de Documents Vidéo basés sur le Formalisme des Graphes Conceptuels pour l'Indexation et la Recherche par le Contenu Sémantique, Ph.D, Thesis, University Joseph Fourier, Grenoble, France, 2005.

[3] Dublin Core Metadata Initiative, Available at: http://dublincore.org/documents/dcmi-terms/

[4] ISO/IEC, Overview of the MPEG-7 Standard (version 8), ISO/IEC JTC1/SC29/WG11/N4980, Klagenfurt, July 2002.

[5] R. Troncy, Nouveaux outils et documents audiovisuels: les innovations du web sémantique, 392 Documentaliste – science de l'information, Vol. 42, n°6, 2005.

[6] N. Hernandez, Ontologie de domaine pour la mosélisation du contexte en recherche d'information, Ph.D. Theses, University Paul Sabatier of Toulouse, France, 2006.

[7] R. Troncy, Formalisme des connaissonces documentaires et des connaissances conceptuelles à l'aide d'ontologie: application à la description de documents audiovisuels, Ph.D. Theses, University Joseph-Fourier, Grenoble, 2004.

[8] A. Isaac, R. Troncy, Conception et utilisation d'ontologies pour l'indexation de documents Audiovisuel, Ph. D. These, École doctorale Concepts et Langages. University Paris IV – Sorbonne, 2005.

[9] C. Desmoulins, M. Grandbastien, Des ontologies pour indexer des documents techniques pour la formation professionnelle, Porceesing of the conférence Ingénierie des connaissances. Toulouse (Centre pour l'UNESCO),2000.

[10] A. Benayache, construction d'une mémoire organisationnelle de formation et évaluation dans un contexte elearning : le projet MEMORAe, Ph. D. Theses, l'UTC. 2005.

[11] M. Buffa, S. Dehors, C. Faron-Zucker, P. Sander, Towards a Corporate Semantic Web Approach in Designing Learning Systems, workshop conference AIED Review of the TRIAL Solution Project, 2005.

[12] A. Bouzeghoub, B. Defude, J. Duitama, C. Lecocq, Un modèle de description sémantique de ressources pédagogiques basé sur une ontologie de domaine» revue sticef 'Sciences et Technologies de l´Information et de la Communication pour l´Éducation et la Formation' Vol. 12, 2005.

[13] A. Hammache, R. Ahmed-Ouamer, Un système de recherche d'information pour le e-learning, Revue Document numérique 1279-5127 - VOL 11/1-2 - 2008 - pp.85-105.

[14] A. Behaz, M. Djoudi, Contribution de génération d'un hypermédia d'enseignement adaptatif à base d'ontologies, 3es Journées Francophones sur les Ontologies JFO Poitiers France, 3-4 Décembre 2009.

[15] A. Dong, H. Li, B. Wang, Ontology-driven annotation and Access of Educational Video Data in E-learning, in E-learning Experiences and Future, Edited by: Safeeullah

Soomro, Publisher: InTech, (pp. 305-326, April 2010, ISBN 978-953-307-092-6).
[16] A. Carbonaro, Ontology-based Video Retrieval in a Semantic-based Learning Environment, Journal of e-Learning and Knowledge Society. Vol. 4, n. 3, September 2008 (pp. 203 - 212).
[17] SMIL : Synchronized Multimedia Integration Langage, Available at: http://www.w3.org/AudioVideo/
[18] OWL Web Ontology Language Overview, W3C Recommendation 10 February 2004. http://www.w3.org/TR/owl-features/ . visited on date 17 mars 2011.
[19] G. Salton, M. Mcgill, Introduction to Modern Information Retrieval», McGraw-Hill, 1983.
[20] J. Martinet, Un modèle vectoriel relationnel de recherche d'information adapté aux images, Ph.D. Theses, Unversity of Joseph Fourier – Grenoble I, 2004.[21] H. Zargayouna, Indexation sémantique de documents XML, Ph.D, Theses, University Paris XI Orsay, 2005.

**Merzougui Ghalia** received a Master in Computer Science from the University of Batna, Algeria, in 2004. She is currently a Professor at the University of Batna, Algeria.
She is a member of (Adaptive Hypermedia in E-learning) research group. She is currently pursuing his doctoral thesis research on the management of multimedia educational content. Her current research interest is in E-Learning, system of information retrieval, ontology, semantic web, authoring and multimedia teaching resource. His teaching interests include computer architecture, software engineering and object-oriented programming, ontology and information retrieval.

**Djoudi Mahieddine** received a PhD in Computer Science from the University of Nancy, France, in 1991. He is currently an Associate Professor at the University of Poitiers, France.
He is a member of SIC (Signal, Images and Communications) Research laboratory. He is also a member of IRMA E-learning research group. His PhD thesis research was in Continuous Speech Recognition. His current research interest is in E-Learning, Mobile Learning, Computer Supported Cooperative Work and Information Literacy. His teaching interests include Programming, Data Bases, Artificial Intelligence and Information & Communication Technology. He started and is involved in many research projects which include many researchers from different Algerian universities.

**Behaz Amel** received a Master in Computer Science from the University of Batna, Algeria, in 2004. She is currently a Professor at the University of Batna, Algeria.
She is a member of (Adaptive Hypermedia in E-learning) research group. She is currently pursuing his doctoral thesis research on the modeling of an adaptive educational hypermedia system. Her current research interest is in E-Learning, Knowledge Engineering, Semantic Web, Ontology, and Learner Modeling. Her teaching interests include Programming, Data Bases, and Web Technology.